\documentclass[conference]{IEEEtran}
\IEEEoverridecommandlockouts

\usepackage[nolist]{acronym}
\begin{acronym}
    \acro{AI}{Artificial Intelligence}
    \acro{API}{Application Programming Interface}
    \acro{ASIC}{Application-Specific Integrated Circuit}
    \acro{ASML}{Advanced Semiconductor Materials Lithography}
    
    \acro{BEOL}{Back-End-Of-Line}
    \acro{BSI}{Bundesamt für Sicherheit in der Informationstechnik}
    
    \acro{CMOS}{Complementary Metal–Oxide–Semiconductor}
    \acro{CPU}{Central Processing Unit}
    \acro{CRT}{Chinese Remainder Theorem}
    
    \acro{DARPA}{Defense Advanced Research Projects Agency}
    \acro{DL}{Deep Learning}
    \acro{DLL}{Dynamic Link Library}
    \acroplural{DLL}[DLLs]{Dynamic Link Libraries}
    \acro{DRM}{Digital Rights Management}
    \acro{DSP}{Digital Signal Processor}
    
    \acro{ECC}{Elliptic-Curve Cryptography}
    \acro{EDA}{Electronic Design Automation}
    \acro{EU}{European Union}
    
    \acro{FEOL}{Front-End-Of-Line}
    \acro{FIPS}{Federal Information Processing Standards}
    \acro{FPGA}{Field-Programmable Gate Array}
    
    \acro{GCD}{Greatest Common Divisor}
    \acro{GDPR}{General Data Protection Regulation}
    \acro{GPU}{Graphical Processing Unit}
    
    \acro{HCI}{Human Computer Interaction}
    \acro{HDL}{Hardware Description Language}
    \acro{HLEGAI}{High Level Expert Group on \ac{AI}}
    
    \acro{IC}{Integrated Circuit}
    \acro{IEC}{International Electrotechnical Commission}
    \acro{IEEE}{Institute of Electrical and Electronics Engineers}
    \acro{IoT}{Internet of Things}
    \acro{IP}{Intellectual Property}
    \acro{IRB}{Institutional Review Board}
    \acro{ISO}{International Organization for Standardization}
    
    \acro{MATE}{Man-at-the-End}
    \acro{MCAS}{Maneuvering Characteristics Augmentation System}
    \acro{MITM}{Man-in-the-Middle}
    \acro{ML}{Machine Learning}
    
    \acro{NIST}{National Institute of Standards and Technology}
    
    \acro{PC}{Personal Computer}
    \acro{PCB}{Printed Circuit Board}
    \acro{PDK}{Process Design Kit}
    \acro{PKI}{Public Key Infrastructure}

    \acro{RE}{Requirements Engineering}
    \acro{RAM}{Random-Access Memory}
    \acro{RTL}{Register Transfer Level}

    \acro{SEM}{Scanning Electron Microscope}
    \acro{SoC}{System on a Chip}
    \acroplural{SoC}[SoCs]{Systems on a Chip}
    
    \acro{TSMC}{Taiwan Semiconductor Manufacturing Company}
    
    \acro{UK}{United Kingdom}
    \acro{USA}{United States of America}
    
    \acro{XAI}{Explainable \acs{AI}}
    \acro{XHW}{Explainable Hardware}
\end{acronym}

\usepackage{cite}
\usepackage{tikz}
\usepackage{tikzsymbols}
\usepackage{amsmath}
\usepackage{calc}
\usepackage{hyperref}
\usepackage{booktabs}
\usepackage{csquotes}
\usepackage{enumerate}
\usepackage{multirow}
\usepackage{rotating}
\usepackage{siunitx}
\usepackage{pdfpages}
\usepackage{enumitem}
\usepackage{xspace}
\usepackage{tabularx}
\usepackage[]{mdframed}
\usepackage{amsthm}

\usepackage[inline,nomargin,index]{fixme} 
\fxsetup{status=draft,theme=color,mode=multiuser,inlineface=\itshape,envface=\itshape} 
\FXRegisterAuthor{ab}{aab}{\color{red}Asia}
\FXRegisterAuthor{js}{ajs}{\color[rgb]{0,0.5,0.16}Julian}
\FXRegisterAuthor{ts}{ats}{\color[rgb]{0.6,0.4,0.8}Timo}
\FXRegisterAuthor{yz}{ayz}{\color{blue}Yixin}
\FXRegisterAuthor{sb}{asb}{\color[rgb]{0.9,0.5,0.8}Steffen}

\newtheorem{definition}{Definition}

\newcommand{\ie}{i.\,e.}
\newcommand{\eg}{e.\,g.}

\newcommand{\etal}{et~al.\@\xspace}

\newcolumntype{x}[1]{>{\centering\let\newline\\\arraybackslash\hspace{0pt}}p{#1}}
\newcolumntype{L}[1]{>{\raggedright\let\newline\\\arraybackslash\hspace{0pt}}m{#1}}
\newcolumntype{C}[1]{>{\centering\let\newline\\\arraybackslash\hspace{0pt}}m{#1}}
\newcolumntype{R}[1]{>{\raggedleft\let\newline\\\arraybackslash\hspace{0pt}}m{#1}}

\newcommand{\participants}{18\xspace}

\usepackage{circledtext}
\circledtextset{boxfill=black}
\circledtextset{boxtype=o}
\circledtextset{charf=\bfseries\Large}
\circledtextset{charcolor=white}
\circledtextset{charshrink=0.6}
\circledtextset{resize=real}
\circledtextset{width=0.8em}

\definecolor{deleted}{RGB}{178,178,178}
\definecolor{inserted}{RGB}{14,130,27}

\newif\ifshowrevision
\showrevisionfalse
\newif\ifshowdeletions
\showdeletionsfalse
\newcommand\del[1]{\ifshowdeletions\textcolor{deleted}{#1}\else\ifhmode\unskip\fi\fi}

\def\orcid#1{\kern.08em\href{https://orcid.org/#1}{\protect\includegraphics[keepaspectratio,width=0.7em]{ORCIDiD_icon.png}}}

\AtBeginDocument{%
  \providecommand\BibTeX{{%
    \normalfont B\kern-0.5em{\scshape i\kern-0.25em b}\kern-0.8em\TeX}}}

\begin{document}

\title{Explainability as a Requirement for Hardware: Introducing Explainable Hardware (XHW)}

\author{\IEEEauthorblockN{
Timo Speith\IEEEauthorrefmark{1},
Julian Speith\IEEEauthorrefmark{2},
Steffen Becker\IEEEauthorrefmark{3}\IEEEauthorrefmark{2},
Yixin Zou\IEEEauthorrefmark{2},
Asia Biega\IEEEauthorrefmark{2},
and Christof Paar\IEEEauthorrefmark{2}}
\IEEEauthorblockA{\IEEEauthorrefmark{1}University of Bayreuth, Bayreuth, Germany}
\IEEEauthorblockA{\IEEEauthorrefmark{2}Max Planck Institute for Security and Privacy, Bochum, Germany}
\IEEEauthorblockA{\IEEEauthorrefmark{3}Ruhr-University Bochum, Bochum, Germany\\Email: timo.speith@uni-bayreuth.de, \{julian.speith, yixin.zou, asia.biega, christof.paar\}@mpi-sp.org, steffen.becker@rub.de}}
\maketitle

\begin{abstract}
In today's age of digital technology, ethical concerns regarding computing systems are increasing. 
While the focus of such concerns currently is on requirements for \emph{software}, this article spotlights the \emph{hardware} domain, specifically \emph{microchips}. 
For example, the opaqueness of modern microchips raises security issues, as malicious actors can manipulate them, jeopardizing system integrity. 
As a consequence, governments invest substantially to facilitate a secure microchip supply chain.
To combat the opaqueness of hardware, this article introduces the concept of \emph{Explainable Hardware}~(XHW). 
Inspired by and building on previous work on Explainable AI~(XAI) and explainable software systems, we develop a framework for achieving XHW comprising relevant stakeholders, requirements they might have concerning hardware, and possible explainability approaches to meet these requirements.
Through an exploratory survey among 18 hardware experts, we showcase applications of the framework and discover potential research gaps.
Our work lays the foundation for future work and structured debates on XHW.
\end{abstract}



\nocite{Speith2024Supplementary}

\section{Introduction}
\label{expl_hw::sec::introduction}

In recent years, researchers have started examining the ethical implications of digital technologies. 
While most work focuses on quality aspects or non-functional requirements (NFRs) of the \emph{software} that runs on these systems \cite{Panesar2019Ethics, Chazette2021Exploring}, the focus of this article is the \emph{hardware}---viz., the microchips---that make these systems run in the first place. 
Consequential initiatives about microchip manufacturing capabilities are currently taking shape around the world, \eg, the European Chips Act~\cite{euchips2022} and the US~CHIPS and Science Act~\cite{uschips2022}.
Through these multi-billion dollar investments, the governments behind them seek to gain more control over the supply of microchips and become less dependent on (untrusted) foreign manufacturers.

A prominent concern is that adversarial (foreign) actors can modify hardware undetected in such a way that the software running on top of it can be manipulated at will. 
Indeed, malicious hardware manipulations can have catastrophic consequences, potentially leading to a complete loss of security or incorrect algorithmic decisions.
Such manipulations include the insertion of a kill switch to render military hardware inoperable under specifiable conditions~\cite{Adee2008Hunt}, manipulation of \ac{ML} accelerators~\cite{Clements2018Hardware, Li2018Hu-Fu, Liu2017Fault}, or the compromising of hardware security primitives~\cite{Becker2013Stealthy}.

The primary problem is that modern microchips are \emph{opaque}.
Microchips have become increasingly complex, culminating, \eg, in the Apple~M1~Ultra with 114~\emph{billion} transistors~\cite{Shankland2022M1}.
Furthermore, non-deterministic design tools automate vast parts of the hardware design process, alienating the designers from the final microchip schematics.
Meanwhile, microchip supply chains are globally distributed and subject to increasing geopolitical tensions~\cite{Mozur2022Eye}, leading to opaque manufacturing processes.
Overall, the resulting opaqueness affects not only downstream stakeholders, such as the end users (\ie, consumers or operators) who interact with the systems but also the experts who design them. 
However, solutions on how to address this opaqueness have not yet been established.

Towards a possible solution, we adapt a prominent concept from discussions on the ethics of other digital technologies: \emph{explainability}. 
In \ac{RE}, the concept of explainability has received a lot of attention recently, and it is rapidly establishing itself as a vital NFR \cite{Chazette2021Exploring, Chazette2020Explainability, Brunotte2022Explainability, Koehl2019Explainability}.

As explainability promises to ease concerns about the security and safety of software and \ac{AI} systems by making them more comprehensible to various stakeholders~\cite{Langer2021What, Chazette2021Exploring}, we argue that adopting the concept of explainability to hardware requirements---thus designing \emph{\ac{XHW}}---has the potential to achieve a similar goal: making hardware more comprehensible to various stakeholders and addressing concerns about security and more.

Based on these considerations, we develop a comprehensive \ac{XHW} framework. Our specific contributions include:
\begin{itemize}
    \item \textbf{Motivating and Defining \ac{XHW}}.
    We argue that hardware is opaque and justify the need for more hardware transparency through, among other things, legislation for trustworthy hardware. 
    As a solution, we propose to transfer the concept of explainability to the hardware domain. 
    We argue that \ac{XHW} is essential to achieve explainability at the system level, building on definitions and models for \ac{XAI}. 
    (\autoref{expl_hw::sec::background} and \autoref{expl_hw::sec::motivation})
    
    \item \textbf{A Framework for \ac{XHW}.}
    We conceive a \emph{framework} for \ac{XHW} encompassing different \emph{stakeholders}, their explainability needs (\ie, requirements and quality aspects related to hardware they are interested in, called \emph{desiderata}), and \emph{approaches} for enhancing the explainability of hardware, drawing on literature from hardware design, manufacturing, and analysis. (\autoref{expl_hw::sec::framework})

    \item \textbf{An Exploratory Study with Hardware Experts.}
    To demonstrate the applicability of our framework, we conduct an exploratory survey among \participants hardware experts.
    Our survey findings hint at distinct needs across stakeholders (\eg, only manufacturers may not care about explainability) and potential limitations of existing approaches (\eg, no approaches are thought to work well for end users). (\autoref{expl_hw::sec::study}) 

    \item \textbf{Potential Applications of the \ac{XHW} Framework.} 
    Finally, we demonstrate how our framework, informed by the results of our study, can help identify directions for future research in \ac{RE}. (\autoref{expl_hw::sec::discussion})
\end{itemize}

\section{Background}
\label{expl_hw::sec::background}

In this work, we refer to a system as something that is usually built and operated by humans and consists of different hardware and software components to serve a specific purpose.
In other words, a system can be anything from a smartphone, a car, to a \acs{PC}. 
A system may also incorporate smaller subsystems, thereby introducing a hierarchy of systems.
We define hardware as physical electronic components, focusing on microchips.
Software is executed on hardware and includes operating systems, user applications, and algorithms like \acs{AI}.

\subsection{Reasons for Hardware Opacity}
Modern microchips are specified and designed using high-level \acp{HDL}.
The resulting schematics are mapped to a technology library that describes all circuit elements available for realizing the design.
This process is known as \emph{synthesis}.
The technology libraries used in this process are typically provided by large  manufacturers, also called \emph{foundries}. 
The implemented design, stripped of all high-level information like hierarchy, labels, and comments, is  passed on to the manufacturer.
They produce the actual chip using the chosen manufacturing technology in one of their production facilities, \ie, a semiconductor manufacturing plant also referred to as \emph{fab}.
In the last step, the fabricated chip can be integrated into various types of systems.

With this background in place, we synthesize the reasons why hardware is opaque by drawing on the three types of opacity in Burrell's work about \ac{ML} algorithms~\cite{Burrell2016How} (see also Mann \etal~\cite{Mann2023Sources}).
First, hardware is opaque to most people due to \emph{technical illiteracy}, even to some experts. 
Their ever-growing complexity makes physical inspection and verification of microchips a challenging task, mastered by only a few highly specialized companies or government agencies worldwide.
Second, modern synthesizers are based on efficient heuristic or even \acs{AI}-based algorithms, which (for complex circuits) leads to different results for every synthesis run.
This correlates to what Burrell calls \emph{opacity characteristic of \ac{ML}}.
Third, developers and manufacturers use obfuscation to protect their \ac{IP} and thus their investments, which relates to opacity as \emph{intentional corporate secrecy}.
The situation is exacerbated by conflicting interests of nation states: they demand openness of foreign hardware while striving to protect domestic systems from external access~\cite{euchips2022, uschips2022}.

\subsection{Towards Hardware Transparency}
The opaqueness of hardware, regardless of its origin, is increasingly recognized as a problem by various stakeholders, as we will outline below. 
In particular, recent research, as well as industry and government initiatives point to the need for a better understanding of hardware to facilitate its transparency.

While the \emph{hardware industry} has historically been seclusive, an open hardware movement has formed in recent years~\cite{Bonvoisin2017Source}, primarily driven by the RISC-V initiative~\cite{openrisc2022}, Google's OpenTitan project~\cite{opentitan2022}, and the CHIPS Alliance~\cite{chipsalliance2022}.
However, these approaches are rarely designed explicitly to foster the understanding of hardware; rather, they tend to cater to vastly different objectives across various stakeholders.

The few studies available to date---although not directly about microchips, but rather about hardware devices in gen\-er\-al---indicate that \emph{end users} have limited understanding of hardware~\cite{Portnoff2015Somebody, Zeng2017End, Lau2018Alexa}.
For instance, end users rarely notice hardware-based webcam LED indicators attached to laptops and smartphones~\cite{Portnoff2015Somebody}; in theory these should serve as privacy safeguards, but in reality they fail to improve users' risk awareness. 
Likewise, end users exhibit limited technical knowledge about smart home devices~\cite{Zeng2017End, Lau2018Alexa}, despite existing research on the information leakage and security vulnerabilities among these devices~\cite{Huang2020IoT}.
Across these examples, a lack of understanding applies to both the software and hardware aspects, jeopardizing end users' privacy and security.

The need for hardware transparency is also increasingly recognized by \emph{governments and policy makers}, making the design and manufacturing of high-end semiconductors a political issue.
For instance, the USA~\cite{uschips2022} and the EU~\cite{euchips2022} have pledged 52 billion USD and 43 billion EUR, respectively, to invest in domestic semiconductor manufacturing~\cite{Clark2021Tech}.
One primary concern with reliance on foreign-made semiconductors is implanted backdoors especially in secure hardware components used by the military or in space~\cite{usvsashoor2010, defscibo2005, euchips2022}. 
To this end, the European Chips Act demands \enquote{a solid understanding of a chip’s architecture}~\cite[p.~44]{euchips2022}.

Altogether, understanding hardware requires efforts from multiple stakeholders as well as dialogues among them. 
What is needed is a comprehensive view of possible directions to achieve hardware understanding for relevant stakeholders. 
This view should not only address technical aspects, but also highlight how and what information must be conveyed to which of the stakeholders involved in the hardware ecosystem.
\section{Hardware Explainability as a Solution for Hardware Opacity}
\label{expl_hw::sec::motivation}
The importance of devising solutions to make hardware understandable to various stakeholders---across diverse contexts and irrespective of the reasons for its opacity---grows.
The objective of explainability aligns perfectly with this goal, aiming to render certain aspects of (\ac{AI}) systems understandable to different stakeholders \cite{Wang2019Designing, Langer2021What, Liao2020Questioning}, irrespective of context or source of opacity \cite{Mann2023Sources}. 
This alignment raises the question: How can explainability be transferred to the hardware domain to make hardware understandable to various stakeholders?

\subsection{From System Explainability to a Definition of Hardware Explainability}
\label{expl_hw::subsec::from_hw_to_system}
As hardware is part of a system, the transfer of explainability to the hardware domain can take place quite straightforwardly.
Recent work has extended explainability from \ac{AI} to software systems \emph{holistically}~\cite{Brunotte2021Welcome, Koehl2019Explainability, Chazette2021Exploring, Brunotte2022Explainability}, as these can also be highly opaque \cite{Mann2023Sources}. 
While the opacity of \ac{AI} may be particularly drastic, other kinds of opacity can already evoke legitimate desires for explainability. 
Chazette \etal~\cite{Chazette2021Exploring} offer a general definition of explainability for software systems:

\begin{mdframed}
    \begin{definition}[System Explainability]
    \label{expl_hw::def::explainability}
    A system $S$ is explainable with respect to an aspect $X$ of $S$ relative to an addressee $A$ in context $C$ if and only if there is an entity $E$ (the explainer) who, by giving a corpus of information $I$ (the explanation of $X$), enables $A$ to understand $X$ of $S$ in $C$. \hfill\cite[p.~200]{Chazette2021Exploring}
    \end{definition}
\end{mdframed}

In this definition, the aspect $X$ to be explained may well be the hardware, since it is a constituent---and, thus, certainly also an aspect---of a software system $S$.
With this finding, a gap in prior research becomes apparent: 
To make a system \emph{truly} explainable, its hardware components must be made explainable as well. 
Merely making an \ac{AI}---or the software that implements it---explainable is insufficient.

To the best of our knowledge, there is no prior research on the explainability of hardware, despite hardware being an essential constituent of systems. 
Furthermore, stakeholders' interests cannot be fully met if we only focus on explainability at the algorithmic level. 
To see why this is the case, let us consider \emph{trustworthiness} as one of the goals of explainability.
Kästner~\etal propose the following operationalization of trustworthiness for software systems~\cite{Kaestner2021Relation}:

\begin{mdframed}
    \begin{definition}[Trustworthiness] \label{expl_hw::def::trustworthiness}
    A system $S$ is \emph{trustworthy} to a stakeholder $H$ in a context $C$ if and only if 
    \begin{enumerate}
        \item[(a)] $S$ works properly in $C$, and
        \item[(b)] $H$ would be justified to believe that (a) if $H$ came to believe that (a). \hfill \cite[p.~171]{Kaestner2021Relation}
    \end{enumerate}
    \end{definition}
\end{mdframed}

For a system to work properly and thus be trustworthy, we must factor in all its constituents---software \emph{and} hardware. 
After all, a system whose hardware does not work properly is hardly worth our trust. 
To find out whether the hardware works properly---and be justified in believing that it does so---we need information on it. 
That is where explainability comes into play \cite{Kaestner2021Relation}: 
by explaining a certain aspect of the system (\eg, its hardware), the understanding of a stakeholder is facilitated, letting them judge whether this aspect works properly~\cite{ Langer2021Auditing}.

\subsection{A Model of Explainability}
\label{expl_hw::subsec::explainability_model}
\begin{figure}[htbp]
    \centering
    \includegraphics[width=0.38\textwidth]{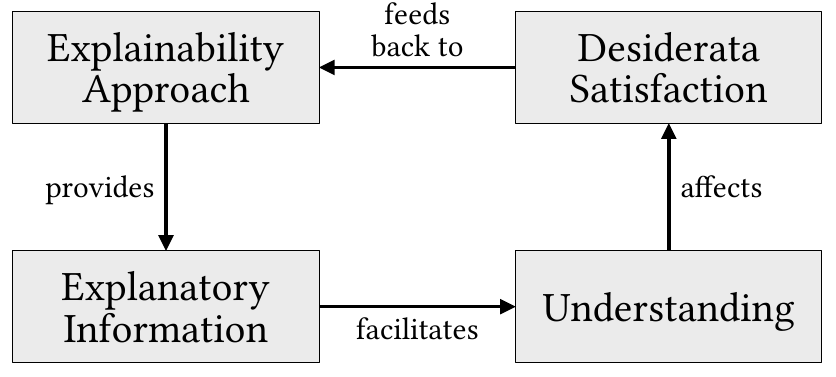}
    \caption{A simplified version of the explainability models that Langer \etal~\cite{Langer2021What} and Hofmann \etal~\cite{Hoffman2018Metrics} have proposed.}
    \label{expl_hw::fig::model}
    \vspace{-1ex}
\end{figure}

Explainability not only promises to help facilitate trustworthiness;
it also supports other desirable quality aspects of systems, such as debuggability, safety, and security \cite{Chazette2021Exploring}. 
These desirable properties are often referred to as \emph{desiderata}~\cite{Langer2021What, Lipton2018Mythos}. 
Coming back to the definition of system explainability (see~\autoref{expl_hw::def::explainability}), all desiderata are \emph{downstream} goals. 
In other words, gaining an understanding of a system through explainability facilitates the satisfaction of a desideratum for a stakeholder. 
Langer \etal~\cite{Langer2021What} and Hoffman \etal~\cite{Hoffman2018Metrics} propose similar models of this process for \ac{XAI} (see \autoref{expl_hw::fig::model}). 

The models propose that explainability approaches (\ie, ways of achieving explainability) provide explanatory information to facilitate a stakeholder's understanding. 
This understanding, in turn, affects the satisfaction of certain desiderata. 
Founded in these models, we develop our \ac{XHW} framework.
\section{A Framework for Hardware Explainability}
\label{expl_hw::sec::framework}
\begin{figure}[htbp]
    \centering
    \includegraphics[width=0.75\linewidth]{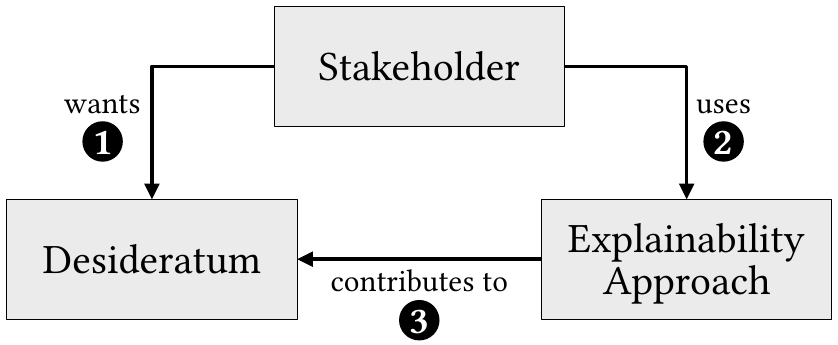}
    \caption{Our \ac{XHW} Framework: Stakeholders want \circledtext{1} desiderata and can use \circledtext{2} explainability approaches that contribute to \circledtext{3} satisfying these desiderata.}
    \label{expl_hw::fig::stakeholder_interactions}
\end{figure}

Through expert knowledge and literature review, we develop our framework for \ac{XHW} in three components: relevant stakeholders (\autoref{expl_hw::subsec::stakeholder}), desiderata (\autoref{expl_hw::subsec::desiderata}), and---in absence of established approaches to \ac{XHW}---existing methods and techniques from hardware design, manufacturing, and analysis that could enhance explainability (\autoref{expl_hw::subsec::current_methods}). \autoref{expl_hw::fig::stakeholder_interactions} shows a high-level illustration of the framework. 

\subsection{Stakeholder Ecosystem}
\label{expl_hw::subsec::stakeholder}
\begin{figure}[htbp]
    \centering
    \includegraphics[width=0.75\linewidth]{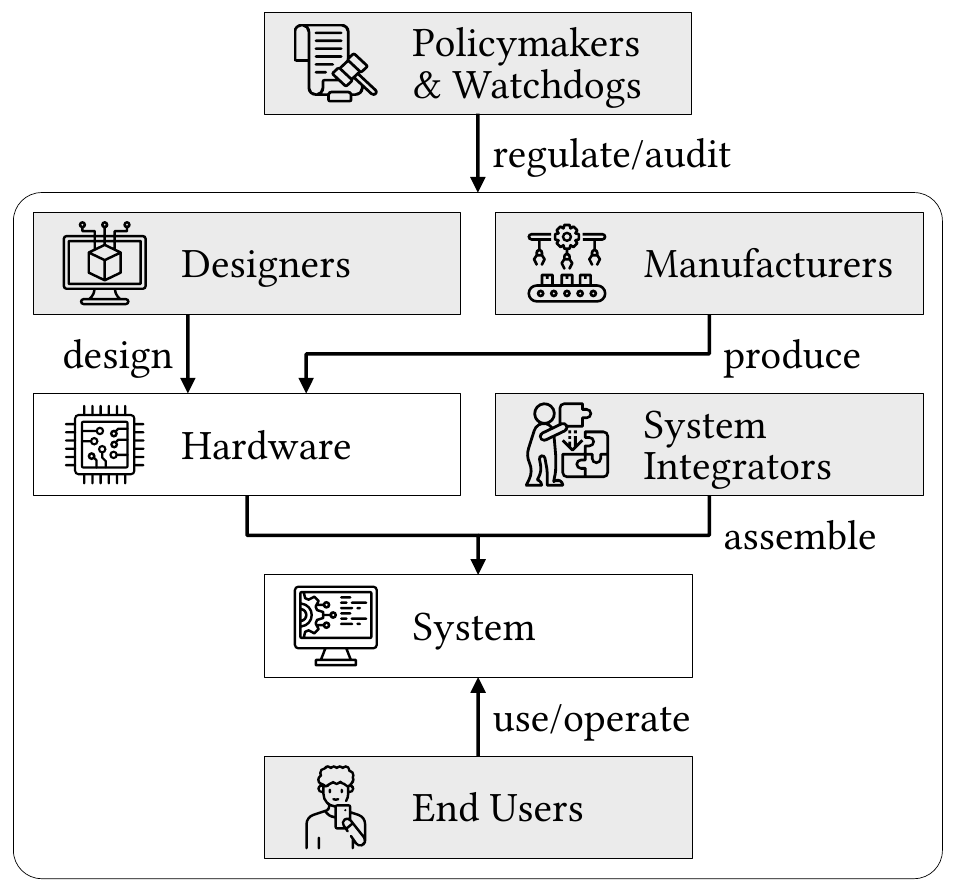}
    \caption{Stakeholder interactions among each other, with the hardware, and with the system as a whole.}
    \label{expl_hw::fig::stakeholder_ecosystem}
\end{figure}

Various stakeholders interact with a hardware-based system throughout its lifetime, each of them having individual desiderata concerning the explainability of hardware.
To address these individual desiderata, we introduce stakeholder categories to classify entities that interact with the hardware by developing, manufacturing, integrating, regulating, or using it. 
These categories are adapted from existing work by Langer~\etal~\cite{Langer2021What} and Tomlinson~\etal~\cite{Tomlinson2022Drivers}: ($i$)~designers, ($ii$)~manufacturers, ($iii$)~system integrators, ($iv$)~policymakers and watchdogs, and ($v$)~end users.\footnote{An entity (a person, company, organization, or government agency) may belong to multiple stakeholder categories at the same time.} 
Below, we outline each stakeholder category (see~\autoref{expl_hw::tab::stakeholder_ecosystem}) as well as the interactions between the stakeholders and the system (see~\autoref{expl_hw::fig::stakeholder_ecosystem}).

\begin{table}[htbp]
    \centering
    \caption{The five stakeholder categories considered in our framework as well as a short description for each of them.}
    \label{expl_hw::tab::stakeholder_ecosystem}
    \begin{tabularx}{\linewidth}{lp{4.75cm}}
        \toprule
        \textbf{Stakeholder} & \textbf{Description}\\
        \midrule
        Designers & Describe the desired hardware functionality in a high-level language. \\
        Manufacturers & Produce the hardware using highly specialized tools and equipment. \\
        System Integrators & Integrate pieces of hardware into a larger system. \\
        Policymakers \& Watchdogs & Set the legal framework in which a system operates and attest adherence thereof. \\
        End Users & Directly use the system or offer a service that is dependent on the system.\\
        \bottomrule
    \end{tabularx}
\end{table}

\subsection{Desiderata}\label{expl_hw::subsec::desiderata}
The goals and purposes of explainability---also referred to as desiderata---depend on the perspectives and requirements of the stakeholders involved. 
Based on these factors, some aspects of explainability may be more relevant than others.  

In \autoref{expl_hw::tab::desiderata}, we give a concise overview of the desiderata most relevant in the context of hardware explainability by adapting existing work on \ac{XAI} and explainable software by Chazette \etal~\cite{Chazette2021Exploring}, Langer~\etal~\cite{Langer2021What}, and Speith \cite{Speith2023Building}.

\begin{table}[htbp]
    \centering
    \caption{The eight desiderata considered in our framework as well as a short description for each of them.}
    \label{expl_hw::tab::desiderata}
    \begin{tabularx}{\linewidth}{lL{6.1cm}}
        \toprule
        \textbf{Desideratum} & \textbf{Description}\\
        \midrule
        Safety & Avoid physical harm to people and the surrounding system. \\
        Accountability & Identify who is responsible in case of failure. \\
        Debuggability & Identify, trace, and correct bugs in order to prevent malfunction. \\
        Legal Compliance & Adhere to the legal framework in which the system operates. \\
        Security & Ensure confidentiality, integrity, and availability of data.\\
        Verifiability & Test correct operations and rule out targeted manipulations. \\
        Trustworthiness & Have correct functionality and be able to demonstrate it.\\
        Trust & Calibrate dependence of a user or component on another component, fitting to its trustworthiness.  \\
        \bottomrule
    \end{tabularx}
\end{table}

\subsection{Leveraging Approaches from Hardware Design, Analysis, and Manufacturing for Explainability}
\label{expl_hw::subsec::current_methods}

The final component of our framework is the approaches that can be leveraged to reach \ac{XHW}.
As we are the first to introduce the notion of \ac{XHW}, there are no established approaches in the literature yet.
Nevertheless, we can adopt approaches from the domains of hardware design, analysis, and manufacturing that contribute to the understanding of hardware and, thereby, improve its explainability according to \autoref{expl_hw::def::explainability}.
We have compiled five such techniques in \autoref{expl_hw::tab::technique_descriptions}. 

\begin{table*}[htbp]
    \centering
    \caption{The five hardware explainability approaches considered in our framework as well as a short description and literature references for each of them.}
    \label{expl_hw::tab::technique_descriptions}
    \begin{tabularx}{\linewidth}{L{2cm}L{5.5cm}L{6cm}c}
        \toprule
        \textbf{Approach} & \textbf{Description} & \textbf{Explainability Benefit} & \textbf{Sources}\\
        \midrule
        Trusted Manufacturing & Retains microchip manufacturing capabilities at domestic facilities to prevent targeted manipulations by a malicious (foreign) fab. & Provides information and assurances about the manufacturing process of a microchip that may not be available about such processes in foreign fabs. & \cite{defscibo2005, Vaidyanathan2014Building, Imeson2013Securing, euchips2022, uschips2022}\\
        Standards \& Certifications & Prescribe a lower bar to be met by microchip components and respective manufacturing processes and verify compliance thereof. & Provide information about the functionality and quality of a microchip, \eg, its use of standardized communication protocols. & \cite{Iso2018RoadVehicles, Brown2000Overview, NIST2019FIPS}\\ 
        Open Hardware & Strives for transparency throughout the supply chain, \eg, with open-source designs, tools, and manufacturing techniques. & Provides information about the architecture, implementation, and manufacturing process of a microchip for others (mostly experts) to inspect. & \cite{opentitan2022, openrisc2022, chipsalliance2022}\\ 
        Testing \& Verification & Evaluates the correctness of a microchip through design and manufacturing using simulation, testing, and formal verification. & Provides information about passed tests and verification procedures of a microchip and thereby also about its functionality and correctness. & \cite{Nayani1998Validation, Geiger1990VLSI, Maly1987Realistic, Kuehlmann1995Verity, Althoff2011Formal}\\
        Physical Analysis & Verifies correctness and checks for undesired information leakage of a microchip using invasive or even destructive physical analysis to rule out malicious modifications. & Provides information about the low-level architecture of a microchip and its regular behavior, as well as its behavior in settings outside the specified operating conditions. & \cite{Torrance2009State, Azriel2021Survey, Kocher1996Timing, Agrawal2002EM, Hsueh1997Fault}\\
        \bottomrule
     \end{tabularx}
\end{table*}
\section{An Exploratory Survey with Hardware Experts}
\label{expl_hw::sec::study}
To showcase the applicability of our framework, we conducted an exploratory online survey with \participants experts from the hardware domain.

\subsection{Survey Method}
\label{expl_hw::subsec::survey}
The primary goal of the survey (see our additional material \cite{Speith2024Supplementary} for the full survey) was to demonstrate that our framework is suitable for identifying hardware explainability gaps (\ie, desiderata that are relevant to a stakeholder but may not be met with the approaches we distilled).
Following the framework proposed in \autoref{expl_hw::fig::stakeholder_interactions}, we asked our participants to \circledtext{1}~determine the relevance of the desiderata for the various stakeholders, \circledtext{2}~identify whether these desiderata can already be achieved using the presented approaches, and \circledtext{3}~how applicable these approaches are to the different stakeholders.

\subsubsection{Study Procedures}
We conducted our survey as an online questionnaire using LimeSurvey.
The median completion time was 19 minutes (min: 9; max: 60; average: 23).
At the beginning, we asked participants for their consent to use the collected data for scientific purposes.
All participants completed the survey voluntarily and received no compensation.

\subsubsection{Participants}
We intentionally targeted our survey to hardware experts, as our primary goal is to test the applicability of our framework, and they have the necessary technical knowledge to help us achieve this goal compared to laypersons.
This recruitment choice reduces the likelihood of misinterpretation and ensures a sufficient baseline of understanding.
Since hardware experts are difficult to reach, we started recruitment in our own professional networks, followed by snowball sampling.
The majority of our \participants participants worked as hardware designers and manufacturers.

\subsubsection{Survey Questions}
After an introduction and explanation of the purpose of our survey, we asked participants for demographic information.
For professional background, we asked participants to indicate their current occupation, professional branch/industry, and field of study.

The core part of our survey consists of three matrix questions in which we asked participants to indicate their perceptions of the relations between stakeholders, desiderata, and explainability approaches. 
We provided definitions of all stakeholders, desiderata, and explainability approaches in the survey to ensure that participants' interpretation coincides with our understanding of the terms.
All three matrix questions asked participants to rate the relationships on a 5-point Likert-like scale. 
The first question asked participants to rate the \emph{importance} of each desideratum for any given stakeholder from \enquote{1 -- not at all important} to \enquote{5 -- extremely important}.
The second question asked participants to rate \emph{applicability} of every explainability approach to each of the stakeholders from \enquote{1 -- not at all applicable} to \enquote{5 -- extremely applicable}.
The third question asked participants to rate the \emph{usefulness} of each approach to satisfy the respective requirement from \enquote{1 -- not at all useful} to \enquote{5 -- extremely useful}.

\subsubsection{Data Analysis}
Since our primary goal of conducting the survey is to demonstrate the usefulness of our framework, we gathered exploratory data on the relationships between stakeholders, desiderata, and explainability approaches. 
To this end, we focused our analysis on descriptive statistics of participants' responses to the three matrix questions.
We calculated the mean values of each cell for the three questions and visualized them as heatmaps (see Figures~\ref{expl_hw::fig::desiderata_to_stakeholder},~\ref{expl_hw::fig::methods_to_stakeholders}, and~\ref{expl_hw::fig::desiderata_to_methods}).

\subsection{Survey Results}
\begin{figure}[htbp]
\centering
    \includegraphics[width=0.7\linewidth]{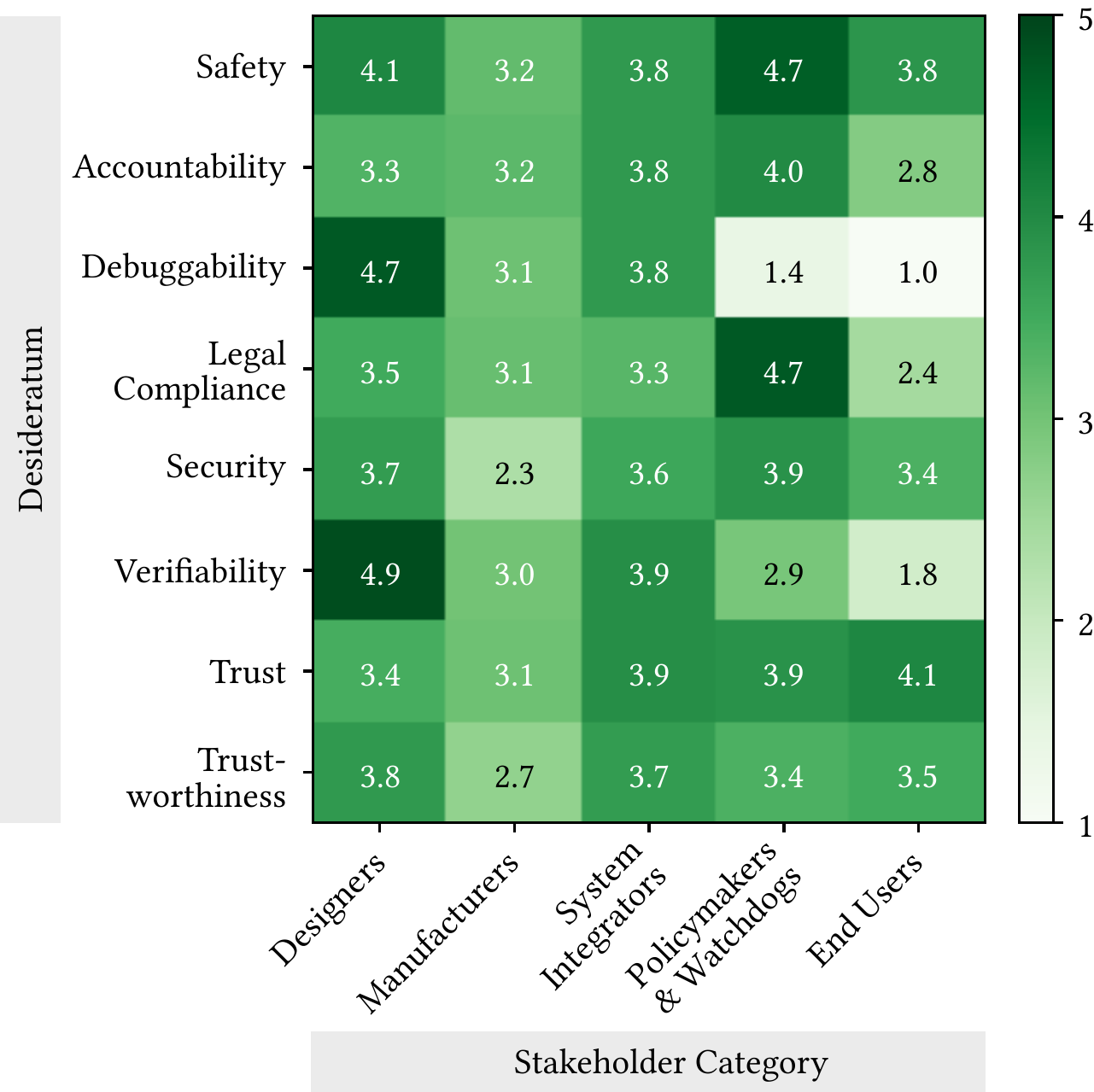}
    \caption{Mean values of participants' responses on the perceived importance of individual desiderata for each stakeholder on a scale from \enquote{1 -- not at all important} to \enquote{5 -- extremely important}.}
    \label{expl_hw::fig::desiderata_to_stakeholder}
\end{figure}

\begin{figure}[htbp]
    \centering
    \includegraphics[width=0.7\linewidth]{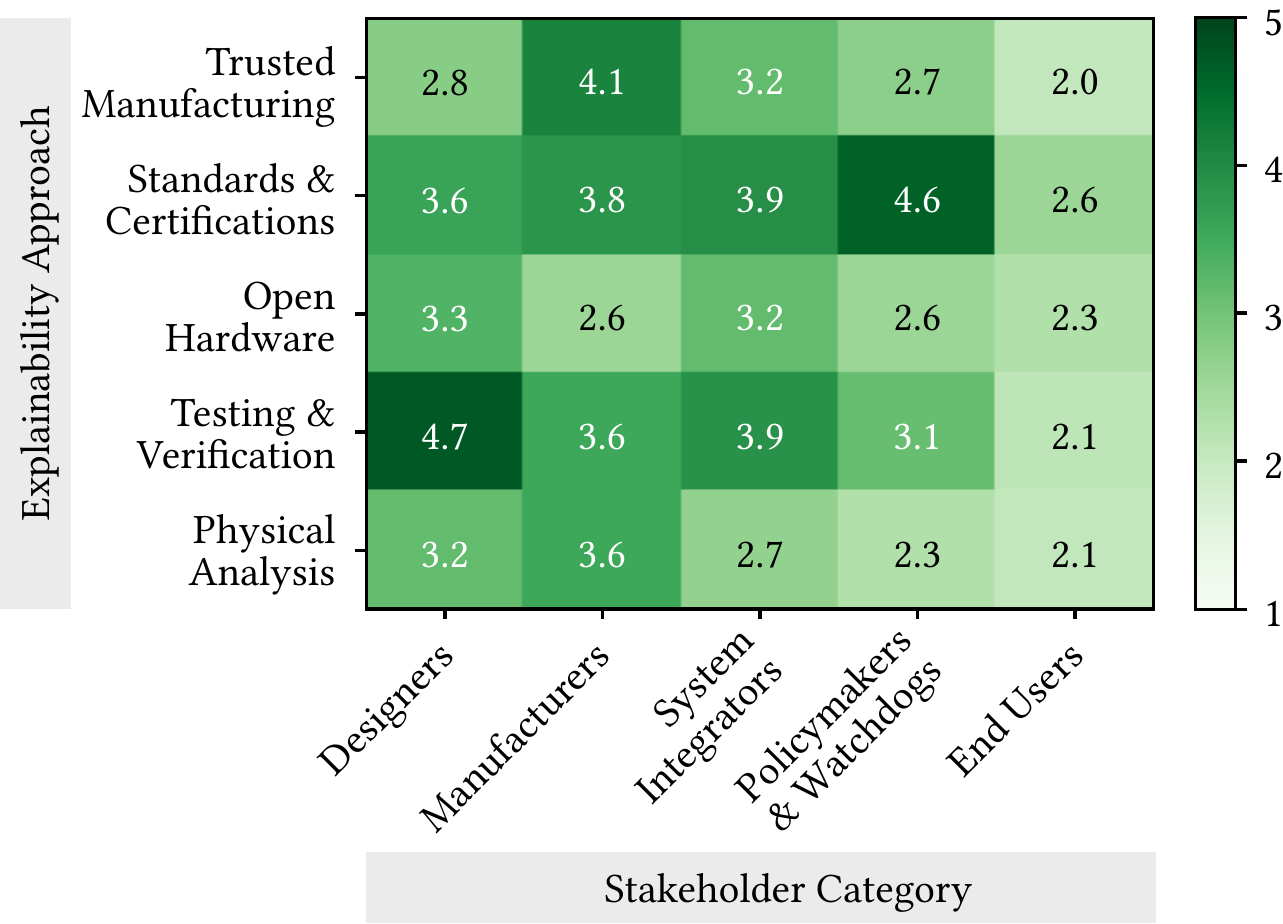}
    \caption{Mean values of participants' responses on the applicability of existing explainability approaches for each stakeholder category on a scale from \enquote{1 -- not at all applicable} to \enquote{5 -- extremely applicable}.}
    \label{expl_hw::fig::methods_to_stakeholders}
\end{figure}

\begin{figure}[htbp]
    \centering
    \includegraphics[width=1\linewidth]{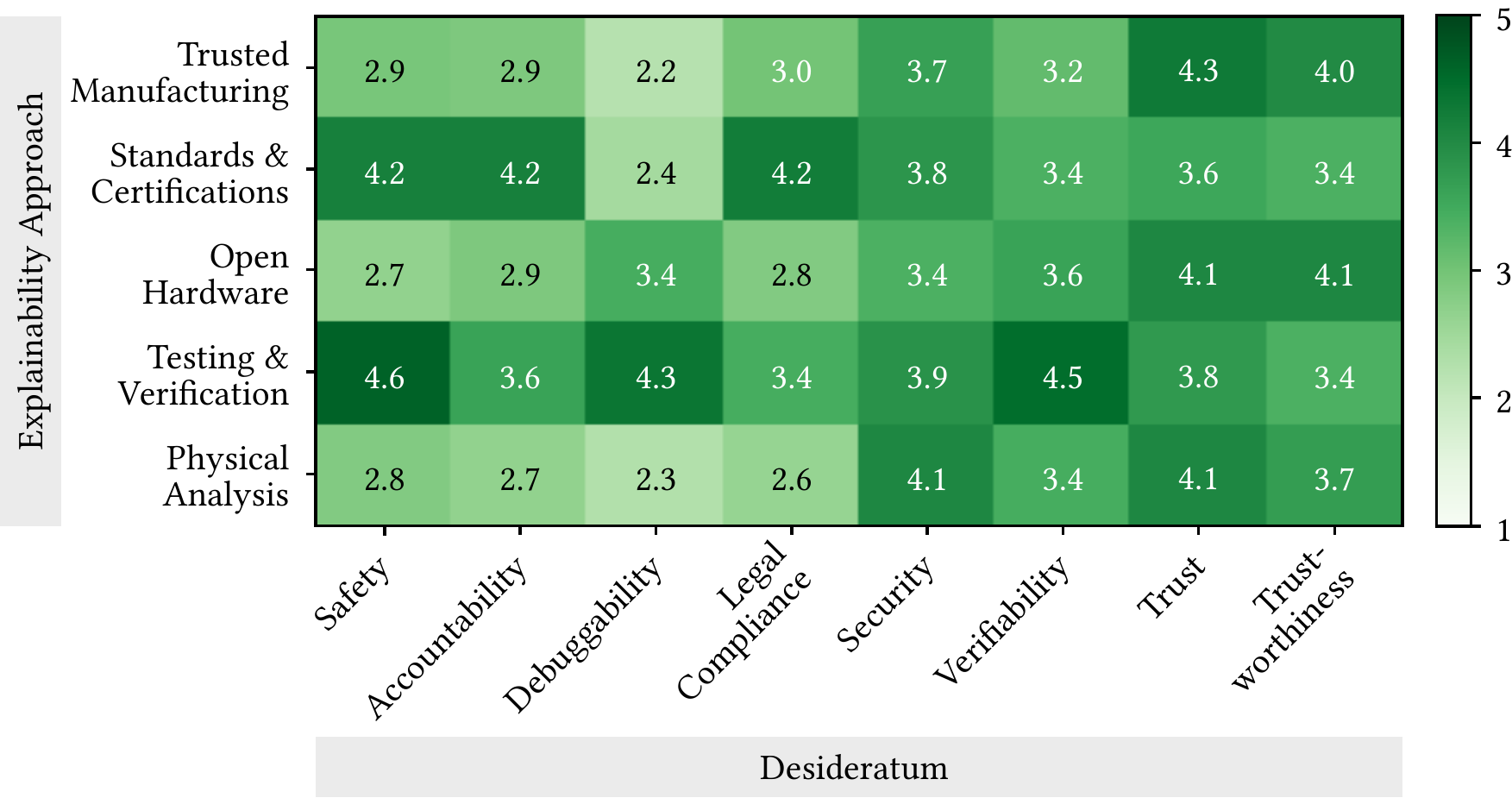}
    \caption{Mean values of participants' responses on the usefulness of existing explainability approaches to satisfy each desideratum on a scale from \enquote{1 -- not at all useful} to \enquote{5 -- extremely useful}.}
    \label{expl_hw::fig::desiderata_to_methods}
\end{figure}

The exploratory survey with hardware experts enables us to uncover potentially satisfied or unmet desiderata for different stakeholders in \ac{XHW}. Specifically, \autoref{expl_hw::fig::desiderata_to_stakeholder} shows the perceived importance of desiderata to stakeholders, \autoref{expl_hw::fig::methods_to_stakeholders} shows the perceived applicability of explainability approaches to stakeholders, and
\autoref{expl_hw::fig::desiderata_to_methods} shows the perceived usefulness of explainability approaches for satisfying desiderata.

Our framework can help guide future research by systematically identifying research gaps. 
We did this through three steps: \circledtext{1}~identify desiderata relevant for a stakeholder, \circledtext{2}~for a given desideratum, determine whether existing approaches can be used to satisfy it, and \circledtext{3}~consider the applicability of the respective approaches to the stakeholder. 
Below, we outline the most salient gaps we identified.

\subsubsection{Low Explainability Needs Among Hardware Manufacturers} 
According to our participants' assessment, manufacturers are not overly concerned with \ac{XHW} (at least not for the desiderata we considered). 
To contextualize the findings, manufacturers do not need to \emph{understand} the hardware they manufacture. 
Instead, they apply quality control measures to verify whether the produced microchips exhibit correct functionality.
Although not directly needed for hardware manufacturing, promoting explainability during manufacturing could help other stakeholders understand the hardware.

\subsubsection{Potential Misalignment of Regulatory Initiatives}
\label{sec:lackalignment}
Open hardware and trusted manufacturing align with and are part of the strategy outlined in the European Chips Act~\cite{euchips2022}.
Indeed, both approaches seem to be particularly well-suited for evoking trust and ensuring trustworthiness.
However, our participants assessed that open hardware and trusted manufacturing have limited applicability for most stakeholders except manufacturers (see \autoref{expl_hw::fig::methods_to_stakeholders}).
This indicates a potential mismatch between regulatory aspirations and the reality.

\subsubsection{Lack of Proper Explainability Approaches for End Users}
\label{sec:lackenduser}
A potential gap for future research emerges as our participants think that the needs of end users are not adequately covered by existing explainability approaches. 
While safety, trust, and trustworthiness are rated as relevant desiderata for end users, there has not been any explainability approach that is at least moderately applicable to them.
Of all the inapplicable approaches, standards and certifications perform best at a mean value of $A_M=2.6$ (see \autoref{expl_hw::fig::methods_to_stakeholders}).
While this approach is certainly not ideal to cover all three relevant desiderata (see \autoref{expl_hw::fig::desiderata_to_methods}), it may still be a good starting point to develop novel techniques that could satisfy end users' needs.

\subsection{Limitations}
As our primary objective with this exploratory survey was to demonstrate that our framework is suitable for identifying potential research gaps, it has some limitations. 
Most importantly, we only had hardware experts as participants, which limits the generalizability of our findings.
We are aware that statements made by hardware experts about end users should be taken with caution, as this is one stakeholder group making statements about a completely different one.
In particular, the potential misalignment of regulatory initiatives (\autoref{sec:lackalignment}) and the lack of proper explainability approaches for end users (\autoref{sec:lackenduser}) may be impacted by this effect. With this exploratory study, however, we primarily aim to demonstrate that our framework is suitable for identifying potential research gaps. Our findings lay the foundation for future work to replicate the same protocol with other stakeholders.

\section{Discussion and Future Work}
\label{expl_hw::sec::discussion}
To date, the understanding of hardware matters has garnered little attention in societal discourse.
However, as becomes evident from media coverage (\eg, \cite{Clark2021Tech, Mozur2022Eye}) and regulatory initiatives (\eg, \cite{euchips2022, uschips2022}), this is likely about to change.
Against this backdrop, the primary goals of our work are to introduce the concept of~\ac{XHW}, motivate why it is needed, and offer a starting point for future research in RE and other fields \cite{Langer2021What} to thoroughly engage with all stakeholders involved. 
This lays the groundwork for requirements engineers of hardware system to formulate \ac{XHW} requirements that suit all stakeholders' needs.

An important direction of future research is to involve laypersons in studies and discourses about hardware.
Such studies could focus on the general \ac{XHW} needs of end users or their understanding of microchips. 
Our framework provides a useful starting point for such investigations.
Moreover, studies with expert populations in the form of interviews could provide the qualitative insights currently lacking, \eg, by probing deeper into the relevance of \ac{XHW} for regulatory initiatives.

Overall, we hope that our framework opens the debate on \ac{XHW} and others will join us in studying the needs of different stakeholders in this new domain more broadly.
Below, we discuss how our work informs concrete directions for future research, noting specific relevance to requirements engineers.

\subsection{Research Direction 1: Filling in Explainability Gaps}
\label{expl_hw::sec::filling}
One potential explainability gap identified through our survey results concerns the end users, as the hardware experts who participated in our study think that none of the proposed \ac{XHW} approaches would apply to end users. 
This does not mean that end users do not need \ac{XHW} at all, as hardware experts speculate that certain desiderata (viz., security, safety, and trust) would still be important to end users.

Nonetheless, research on end users' understanding of hard\-ware is extremely limited. 
While it is reasonable to expect that end users have little understanding of hardware due to a lack of technical expertise, it is important to uncover their specific mental models of how hardware works. 
Drawing from work on end users' mental models of other topics such as computer security~\cite{wash2010folk} and the internet~\cite{kang2015my}, we expect that research on end users' mental models of hardware could shed light on the information needed in explaining hardware to end users.

Based on these deliberations, we derive our first two research questions (RQs) for future work:

\begin{mdframed}
    \textbf{RQ1:} What are end users' mental models of hardware?\\
    \textbf{RQ2:} What information about hardware is required to align end users' mental models with system models?
\end{mdframed}

In order to answer RQ1, a qualitative user study could elicit end users' mental models of hardware.
For RQ2, we would need to verify, again in a user study, whether the approaches of our framework are really insufficient to produce information that can reconcile end users' mental model with the actual system models.
If the results for RQ2 show that the existing \ac{XHW} approaches are indeed insufficient for end users, new approaches must be developed for their needs.  

Research on usable security and privacy has been successful in developing and advocating standardized labels for \acs{IoT} devices~\cite{Emami-Naeini2020Ask} and smartphone apps~\cite{Cranor2022Mobile}, drawing inspiration from food nutrition labels. 
Against this background, we imagine that hardware labels could function as an explainability approach for hardware that is particularly suitable for end users. 

Accordingly, we formulate a third RQ for future research:

\begin{mdframed}
    \textbf{RQ3:} What information shall hardware labels provide to meet the \ac{XHW} requirements and needs of end users?
\end{mdframed}

There are already standards and norms in the hardware community that requirements engineers can draw from to answer this RQ, such as \acs{ISO}~26262~\cite{Iso2018RoadVehicles} and \acs{IEC}~61508~\cite{Brown2000Overview} for hardware safety and \acs{FIPS}~PUB~140-3~\cite{NIST2019FIPS} for hardware security.
A comprehensive requirements elicitation and interpretation, as well as validation (\eg, through prototyping), would be next steps here.

\subsection{Research Direction 2: Devising New XHW Approaches}
We see the potential of adapting existing research on \ac{XAI} (and explainability in general) to generate new approaches for \ac{XHW}. 
In line with the above idea of hardware labels, research on model cards \cite{Mitchell2019Model} and datasheets for datasets \cite{Gebru2021Datasheets} can already be found in the debates on \ac{XAI}. 
This research could serve as a basis for adapting results from \ac{XAI} and, in particular, \ac{XAI} approaches for \ac{XHW}.

Further \ac{XAI} approaches we envision to be adaptable are~\emph{counterfactual explanations} \cite{Wachter2017Counterfactual} and~\emph{LIME} \cite{Ribeiro2016Why}. 
Counterfactual explanations allow system integrators to check whether a given microchip conforms to its original schematic by examining what needs to change in the microchip's input for a given deviation in its output.
Employing the idea behind LIME, hardware designers could create surrogate models of a microchip for specific input regions by varying its input and looking at the changes in the output. 
These models could be used to simulate and check the microchip's behavior.

Against this background, another avenue for future research would be to analyze the extent to which (ideas behind) \ac{XAI} approaches can be transferred to \ac{XHW} to devise new explainability approaches for hardware, leading to the following RQ:

\begin{mdframed}
    \textbf{RQ4:} How can \ac{XAI} approaches be adopted for \ac{XHW}?
\end{mdframed}

A possible starting point for answering this RQ would be to distill features of \ac{XAI} approaches and see whether these are useful for hardware. 
Such features could be taken from \ac{XAI} taxonomies or reviews (\eg, \cite{Speith2022Review, Langer2021What, Chazette2021Exploring}), and their suitability for \ac{XHW} could be discussed in a workshop with hardware experts.

\subsection{Research Direction 3: The Right to Repair} 
Given the considerable ecological footprint of producing modern electronics, repairability has become a concern for both legislators and end users. 
In fact, a \emph{right to repair} is enshrined in law in more and more countries (\eg, the United States, the United Kingdom, and India), requiring, among others, that a device should be constructed and designed in a manner that allows repairs to be made easily (see, e.g., \cite{svensson2018emerging}).
Providing details on microchip functionality and interfaces, which aligns with the goals of \ac{XHW}, could empower end users to properly judge causes for microchip breakdown, determine the suitability of replacement parts, and learn how to replace affected components. 
Furthermore, information on utilized materials and compositions could aid recycling efforts of products broken beyond repair.

Thus, the final avenue for future research that we want to emphasize is to find out if and how \ac{XHW} could support a right to repair. 
We derive our last RQ for future research:

\begin{mdframed}
    \textbf{RQ5:} How can \ac{XHW} help to provide and obtain the information necessary to exercise the right to repair?
\end{mdframed}

The hardware labels mentioned above (see \autoref{expl_hw::sec::filling}) could be a first step to answer this RQ, requiring future research based on our framework.
Overall, however, a thorough requirements elicitation and interpretation, especially from system integrators, would be valuable in answering this RQ. 
The collected requirements would then have to be reviewed by legal scholars to determine whether they cover the law.
\section{Conclusion}
\label{expl_hw::sec::conclusion}
In this article, we propose \acf{XHW} as a new concept and motivate its necessity through legislative initiatives and existing findings on \ac{XAI} as well as software explainability. 
Motivated by these findings, we conceive a framework comprising different stakeholders, their desiderata, and existing approaches towards \ac{XHW}.
In an exploratory survey among 18 hardware experts, we demonstrated its applicability and usefulness to identify research gaps.

Our framework paves the way for further studies, \eg, to explore end users' needs concerning \ac{XHW} more concretely.
In particular, it is plausible that end users will interact differently with a system, depending on the installed hardware and their level of information about it.
In this line of thought, prospective \ac{XHW} approaches could go hand in hand with right-to-repair legislation to facilitate repair and recycling.
Inspirations for such approaches could be taken from \ac{XAI}, \eg, in the form of labels, or by adopting other existing \ac{XAI} techniques.
Overall, \ac{XHW} represents a vital new research direction to which \ac{RE} researchers can contribute significantly.

\section*{Acknowledgments}
We would like to thank 
Gabriel Lima and  all participants of the EIS colloquium for feedback on early versions of this article. Furthermore, we would like to thank three anonymous reviewers and Hannah Deters for helpful comments on a more advanced version of this article. Special thanks go to Jakob Droste and Markus Langer for feedback on all versions.

Work on this paper was funded by the Deutsche Forschungsgemeinschaft (DFG, German Research Foundation) under Germany's Excellence Strategy---\href{https://casa.rub.de}{EXC 2092 CASA}---390781972, through the DFG grant 389792660 as part of \href{https://perspicuous-computing.science}{TRR~248}, and by the Volkswagen Foundation grants AZ~9B830, AZ~98509, and AZ~98514 \href{https://explainable-intelligent.systems}{\enquote{Explainable Intelligent Systems}} (EIS). 

The Volkswagen Foundation and the DFG had no role in preparation, review, or approval of the manuscript; or the decision to submit the manuscript for publication. 
The authors declare no other financial interests.

\bibliographystyle{IEEEtran}
\bibliography{bibliography}


\end{document}